# Influence of boundary conditions on 2-fluid Systems under horizontal vibration

## P. Evesque

Lab MSSMat,  UMR 8579 CNRS, Ecole Centrale Paris
92295 CHATENAY-MALABRY, France, e-mail: evesque@mssmat.ecp.fr

### Abstract:

*This paper is concerned with 2-phase systems under vibration in gravity condition, when the gravity is perpendicular to the direction of vibration. It tries and demonstrates that, even in such a restricted case, the patterns which can be formed are very sensitive to the cell shape, to the boundary conditions and to the direction and the mode of vibration, i.e. linear, rotational,…*

**Pacs # :** 5.40 ; 45.70 ; 62.20 ; 83.70.Fn

______________________________________________________________________

In some case, vibration on liquid-sand systems generate a solid wavy relief at the surface of sand; this relief does not disappear when vibration is stopped proving the solid-like nature of sand (*cf.* experiment #1 below). In some other cases which looks like similar, this pattern does not seem to be observed (*cf.* experiment #2 below) ; instead, vibration generates a frozen wavy pattern at lager amplitude; this "static" relief disappears when vibration is stopped proving the liquid like nature of the sand. Are these results compatible? Why do similar experimental conditions lead to different behaviour? The aim of this paper is just to discuss this point which has been arisen during a meeting on miscible fluids :

Indeed, vibration-induced flows are commonly observed in liquids, even when these liquids are filling up the container that is vibrated. One of the main reason is momentum preservation ; but the mode of vibration, *i.e.* translation, rotation,…, is also strongly important. An other important parameter is the boundary conditions. This is just what it is wanted to emphasise in this paper since stating correctly the problem shall help controlling experimental conditions and reducing artefacts when vibrations play some important part ; in particular, this happens for experiments in weightlessness condition.

So, we describe first two vibration experiments on 2-phase systems that seem to be quite similar, but that lead to two different behaviours. We show that their boundary conditions are quite different, which explains their difference of behaviour. We generalise the approach and propose a set-up which allows to pass from the first behaviour to the other one.

## 1.  The experiments :

### *1.1.  Experiment #1 , which uses rotation vibration :*

We start with the experiment on a 2-phase system, *i.e.* liquid-&-sand, proposed by Stegner and Wesfreid, or by Scherer *et al.* [1]. In their paper these authors use an annular cell of rectangular section (height h, mean radius $R_t$, typical diameter of the section $\Delta R=2R_s$) on which cyclic rotation is applied (angular amplitude $\delta\theta=L/(2R_t)$, frequency f) whose direction is parallel to the annular-cell axis and to gravity. One observes after a while, and above a threshold, the formation of a periodic relief made of ripples. These ripples form slowly at the sand surface as soon as the reduced acceleration $\gamma/g = 4\pi^2 f^2 \delta\theta R_t/g$ is large enough. Their final wavelength is proportional to the vibration amplitude L, *i.e.* $\lambda \approx 2/3 L$; the final slope of the relief is found constant, equal to 0.37 about. In general, the acceleration is kept small $\gamma/g<0.2-0.5$, otherwise other effects occur due to partial liquefaction of sand. Indeed, as the relief remains when vibration is stopped, it shows that deformation of **sand** obeys a plastic-like (or **solid-like**) behaviour under these experimental conditions.

In order to estimate the threshold value $\gamma_m/g$ of the reduced acceleration $\gamma/g$ above which sand liquefaction may perturb the ripple formation, one can proceed as follows: be $\Phi$ the sand porosity, be $\rho_s$, $\rho_l$, $\varrho$ the densities of a sand grain, of liquid and of the average medium respectively, *i.e.* $\varrho= \rho_l\Phi+ \rho_s(1-\Phi)$ or $(\varrho-\rho_l)= (\rho_s-\rho_l)(1-\Phi)$; consider a horizontal slice of height $\delta h$ of sand and be S its horizontal surface; as sand is filled with liquid, this slice is submitted to an inertial force $F_i= \gamma \varrho \, \delta h S$ and to a friction force $F_f$ which shall be smaller than $F_{fm}=\tan(\varphi)(\rho_s-\rho_l)(1-\Phi)g \, \delta h S=\tan(\varphi)(\varrho-\rho_l)g \, \delta h S$, with $\tan(\varphi)\approx 0.55$. Static equilibrium, and hence no liquefaction condition, is obtained if $F_i<F_{fm}$, which leads to $\gamma_m/g= \tan(\varphi)(1-\Phi)(\rho_s-\rho_l)/\varrho$, *i.e.* $\gamma_m/g= \tan(\varphi)(\varrho-\rho_l)/\varrho$.

So, under these circumstances $\gamma<\gamma_m$, the sand bed is immovable in the cell frame; but the water bed, which is on top of it, moves periodically in this frame due to inertia. It is this periodic motion which forces the generation of the periodic relief: let consider the case when the thickness $\delta=[\nu/(\pi f)]^{1/2}$ of the viscous boundary layer is small compared to h and to $\Delta R$, but large compared to the typical grain size d, $\nu$ being the kinematic viscosity of the liquid. This periodic flow generates an important friction which forces the sediment to go to and fro at the interface. This motion allows to destabilise the flat interface above a given threshold and the growth of the relief [2]. Indeed, as soon as the relief starts, non linearity in the Navier-Stokes equation allows to generate a mean convection flow from the periodic forcing as in the classical Schlichting and acoustic streaming mechanisms [3]; this mean flow enhances the relief [2], which enhances the flow,... And the solid wavy relief builds up.

## *1.2. Experiment #2, which uses linear translation vibration:*

On the other hand, in [4] Ivanova *et al.* have used linear vibration (amplitude A, frequency f) on similar 2-phase systems contained in cylindrical or rectangular closed cells. They did not report having observed the ripple formation reported in [1]; on the contrary they claim that an other kind of standing relief is generated at much larger vibration amplitude ; this relief grows up much faster than the one in [1]. It is also steady in the cell frame (as in [1]), but it disappears as soon as the vibration is stopped. Owing to this, the **sand** behaviour shall be considered as **liquid-like**. Furthermore, the amplitude of vibration for which this relief is observed corresponds always to $\gamma/g > \tan(\varphi)$, where $\tan(\varphi)$ is the solid friction of sand ; and the real parameter that controls the relief has been found to be $(Af)^2/g$ .

For this reason, the relief formation is attributed to a Kelvin-Helmholtz-type mechanism that assumes that any flat interface separating two liquids in relative motion is unstable due to the destabilising effect of Bernoulli pressure on interface disturbance. This mechanism is similar to the mechanism of the formation of waves by wind on the sea, the role of wind is plaid by the top fluid here. Also, the relief does not propagate here because the "wind" flows alternatively in one way and in the other way, so that the

This physical explanation may look simple ; anyway, Lyubimov and Cherepanov [5] have proposed within this scheme a complete stability analysis of a flat interface in the case of two non viscous incompressible liquid with surface tension. This approach was partly extended to the liquid-sand case [6] and explains correctly the experimental data [4].

## 2. The Question :

Are these two results incompatible? In order to demonstrate their compatibility, it is worth to emphasise their difference and to demonstrate that the flows, which are generated in these two different experiments, are due to two different mechanisms.

• *The explanation of experiment #1* is based on (i) a solid-like behaviour of the sand in the cell frame, while liquid is immovable (due to inertia in the Galilean frame). So liquid flows periodically above the sand in the cell frame. This allows the mechanisms of ripple formation to start forming ripples at the surface of sand. However, as soon as vibration amplitude generates an acceleration stronger than g $\tan(\varphi)$ both liquid and sand flow in the cell frame, at least at some phase of the period for sand ; this liquefies the sand surface which flattens.

At much larger amplitude of vibration the sand is completely fluidised and flows by inertia in the same direction as the liquid in the cell frame. Or what is the same, as soon as friction effect can be neglected, the two "liquids" are immovable in any Galilean frame in which the tore centre is fix. So, they *"move"* at the same speed and no Kelvin-Helmholtz-type instability can then develop.

As we will see now, the main difference between experiment #1 and experiment #2 comes from boundary condition since the cell is closed in experiment #2 while it allows continuous flow line of toroidal shape in experiment #1. This changes the characteristics of the flow.

• *The explanation of experiment #2* is based on the two following points : (i) there is no liquid motion when the sand has a solid-like behaviour, *i.e.* $\gamma < g \tan(\varphi)$ (or better $\gamma/g <$ tan($\varphi$) ($\rho$-$\rho_l$)/$\rho$), because the cell is closed ; (ii) as soon as the sand is liquefied, *i.e.* $\gamma > g$ tan($\varphi$), vibration forces the liquid and the sand to perform a relative motion compared to each other, because the two phases have different densities. This relative motion is characterised by its typical relative speed $v_r$, which depends on the vibration parameter Af and on the density of the two phases, $\rho_1$, $\rho_2$. This forces the generation of new effects ; in particular it forces the generation of an instability of the Kelvin-Helmholtz-type as soon as : (i) $\gamma >> g$ , (ii) the frequency f is large enough and (iii) the relative speed $v_r$ overpasses a given threshold.

## 3. Demonstration of the influence of the boundary conditions :

So one sees that conditions of flow generation are quite different in experiments #1 & #2 . In order to exemplify the difference between the two experiments, let us now consider two new experiments that are quite similar to the previous ones, but with slight differences in boundary conditions :

• ***In experiment #1b***, which is quite similar to experiment #1, a tore cell is used and is excited by a periodic rotation whose axis is the tore axis ; it is filled up with the 2-phase system of sand and liquid. However, the cell is now modified in order to forbid circular flow line ; this is obtained by building a wall located on one section of the tore. This breaks then the toroidal symmetry. Under these experimental conditions, no flow is generated in the liquid when the sand has a solid-like behaviour, *i.e.* when $\gamma < g$ tan($\varphi$) (or better $\gamma_c/g =$ tan($\varphi$) ($\rho$-$\rho_l$)/$\rho$) ; this shall inhibit the mechanisms of ripple formation as it was observed in experiment #1. On the contrary, the two phases are now forced to perform relative motion due to inertia, when sand is liquefied, *i.e.* when $\gamma > g$ tan($\varphi$) ; this allows the generation of a "frozen" relief by a mechanism of Kelvin-Helmholtz type at the interface of the two liquids, when vibration parameter $v_r = \Delta\theta$ f is large enough. This experiment has been already performed [8]. Slight modification has to be introduced in order to take into account the effect of the centrifugal force with the rotational vibration, which inclines the interface compared to the horizontal.

      One shall note also that more complicated effect can take place in the case when particles are larger than the boundary layer thickness $\delta$, and when the axis of rotation is not vertical. Indeed, in this case the relative motion of the fluid and of the particles generates new forces, which are a combination of centrifuge- and of Coriolis- forces. These new forces attract the dense particles to the centre of rotation, so that the dense particle can levitate above a given threshold [9].

• ***In experiment #2b***, a horizontal rectangular cell is used and is subject to linear vibration as in experiment #1 ; however, it is no more closed at the two ends, but its two ends are connected to two large containers via two elastic (deformable) tubes. In this case as soon as vibration frequency is large enough to allow to neglect viscous effect the liquid flows freely in the pipe even when the sand has a solid-like behaviour, *i.e.* when $\gamma < g$ tan($\varphi$). So, under such circumstances, *i.e.* $\gamma < g$ tan($\varphi$), the liquid is immobile in the lab frame, but the sand follows the pipe motion ; this allows the mechanisms of ripple formation to take place, which builds up a solid-like wavy relief. However, when vibration amplitude is increased and when acceleration γ overpasses the value $\gamma_c$=g tan(φ) (or better $\gamma_c$/g= tan(φ) (ρ-$ρ_l$)/ρ), the sand is liquefied at some phase of the period. At larger vibration amplitude, and when frequency f is large enough to allow neglecting friction effects in sand- and liquid- phases, both phases become immobile in the lab frame, so that they do not exhibit any relative motion anymore. So one can not observe an instability of the Kelvin-Helmholtz type developing in experiment #2b at large amplitude of vibration.

As a matter of fact, it is worth noting that results of experiment #1b (*resp.* #2b) are similar to those of experiment #2 (*resp.* #1). Indeed, experiment #1b (*resp.* #1) is just the experiment #2 (*resp.* #2b) in the limit of an infinite tore radius $R_t$ , keeping constant the vibration amplitude L=A=$R_t$ Δθ. So, this example demonstrates and illustrates the importance of the boundary conditions and of the polarisation of periodic motion in experiments using fluids and vibrations.

This remark allows settling a model that allows passing continuously from experiment #2 to experiment #1 :

## 4. Generalisation : Theoretical modelling

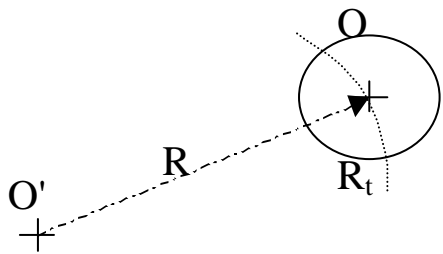

Indeed, to do so, one shall note first that tore-like flow lines are generated by inertia in any closed cell filled with liquid that is started rotating. So, let us consider a cylindrical finite closed cell of radius $R_t$ , centred on point O, on which a periodic motion is imposed ; let us fill it up with liquid and sand and let us also assume that the periodic motion consists of a periodic rotation around a point O' which might be different from O . Let us note R the distance OO'.

So, when R<<$R_t$, the experimental condition looks very much like the one of experiment #1 while they look very much like the ones of experiment #2 when R>>$R_t$ .

We will now study what shall be observed in such a cell under such a periodic motion. However, in order to simplify the modelling we consider now a tore-like cell centred on O instead of the previous cylindrical one. Defining $R_t$ as the tore radius and $R_s$ the radius of the section of the tore, we assume also $R_s$<<$R_t$ for convenience. We will now describe what shall be observed in the different cases of (i) a single solid-like phase filling up the cell, (ii) a single liquid-like phase, and (iii) a 2-phase system filling up the cell.

• *Solid-like phase:* Be R Δθ sin(2πft), the motion of the tore centre, with Δθ the rotation amplitude and A=RΔθ the amplitude of motion of the tore centre. The motion δM of any point M of the tore is given by:

$$\delta M = \delta O + OM \wedge \Omega$$

with $\Omega = 2\pi f \, \Delta\theta \cos(2\pi ft)$. $\delta M$ is also the displacement of any point of a solid-like phase that is glued to the wall of the cell so that it is immobile in the tore frame. Such a solid-like phase can be made up of sand if $\gamma < g \tan(\varphi)$ or of a liquid with a large kinematic viscosity $\nu$, if the frequency is small enough so that the boundary layer extends over the whole sample; this requires $\delta = [\nu/(\pi f)]^{1/2} >> R_s$. Two typical amplitudes of motion and of speed shall be defined. The first one shall refer to the motion of the tore centre, *i.e.* $A_o = R \, \Delta\theta$ and $v_o \approx 2\pi f \, R \, \Delta\theta$. The other one corresponds to the variation of these amplitudes on a distance equal to a radius $R_t$ of the tore, *i.e.* $A_t = R_t \, \Delta\theta$ and $v_t \approx 2\pi f \, R_t \, \Delta\theta$.

- ***Homogeneous perfect liquid phase:*** However, as soon as frequency becomes large enough so that $\delta = [\nu/(\pi f)]^{1/2} << R_s$, inertia plays the main role in the largest part of the liquid phase. This part follows a motion which is different from the solid tore since conservation of momentum tends to impose to it the following motion:

$$\delta M = \delta O$$

In the cell frame, the amplitude of motion of the perfect liquid is then $R_t \, \Delta\theta$ and its typical speed variation is $v_r \approx 2\pi f \, R_t \, \Delta\theta$, which has to be compared to the typical amplitude $A_o = R \, \Delta\theta$ and speed $v_o \approx 2\pi f \, R \, \Delta\theta$ of the tore centre.

A similar motion is obtained for a liquefied sand phase, *i.e.* when and $\gamma >> g \tan(\varphi)$.

- ***2-phase system:*** When the system filling the tore is composed of two different phases in half-half proportion, different cases appear depending of the nature of the phases:

  - If the two phases are solid-like they move as a whole with the tore; they have the same speed and amplitude defined by $(A_o = R \, \Delta\theta, v_o \approx 2\pi f \, R \, \Delta\theta)$ and $(A_t = R_t \, \Delta\theta, v_t \approx 2\pi f \, R_t \, \Delta\theta)$.

  - If one of the phase is solid-like and the other one liquid-like, the first one is immovable in the tore frame, while the second moves; so if the solid-like phase can be deformed plastically, the mechanism of experiment #1 can act and deforms the interface plastically. The typical speed that has to be considered is the relative speed $v_r \approx 2\pi f \, R_t \, \Delta\theta$, which can be much smaller than the vibration speed of the tore, *i.e.* $v_o \approx 2\pi f \, R \, \Delta\theta$, if $R_t << R$.

  - Consider now a system of two liquid-like phases filling up the tore and consider the case when their boundary layer thickness $\delta_1$ and $\delta_2$ are much smaller than $R_t$. The part of the two phases that moves with the tore is then negligible. Consider first the role plaid by momentum preservation as exemplified by experiment #1. Indeed, momentum preservation imposes that the two phases move both with approximately the same speed $v_r \approx 2\pi f \, R_t \, \Delta\theta$ in the tore frame.

So, the mechanism of momentum preservation invoked in experiment #1 cannot generate a static relief generation of the Kelvin-Helmholtz type, at the interface of two (perfect) liquids, nor a wavy solid relief due to some Schlichting mechanism.

However mechanism of experiment #2 still applies there, if gravity is perpendicular to the vibration. The typical relative speed $v_i$ imposed by inertia to the two liquids is $v_i \approx (\rho_1-\rho_2)/(\rho_1+\rho_2) v_o$, with $v_o \gg 2\pi f R \Delta\theta$. This relative motion can generate a frozen relief due to a mechanism of the Kelvin-Helmholtz type above a given threshold of excitation, as in experiment #2 ; this threshold depends on the surface tension ; it can be zero if capillary forces are null. However, in that case, the relief wavelength and amplitude tend to zero with the excitation ; this shall be the case in the case of liquid-sand mixture.

Furthermore, in the case of liquid-sand mixtures, at least two other limitations are required : (i) frequency of vibration shall be small enough so that the boundary layer thickness $\delta = [\nu/(f\pi)]^{1/2} > d_o$ be larger than the pore size $d_o$. Indeed, this is a needed condition, since it forces the liquid contained in the pores to move coherently with the sand itself ; hence, it maintains the coherence of the global phase. (ii) The wavelength $\lambda$ of the relief shall be larger than the grain size, otherwise the relief is not observable.

## 5. Conclusion : Application to space experiment

It is wellknown that vibrations perturb strongly experiments with fluids in micro-gravity environment. This is mainly due to the fact that the experiments which are performed are often sensitive to gravity and hence to inertia. In this case, vibrations induce flows always in inhomogeneous fluids or in homogeneous fluids when rotation is applied ; these flows act on the fluid interfaces, perturb their shape and modify the position of the fluids ; these mechanisms become quite important because there is no more any gravity field which stabilises the interface shapes and positions [10] ; so these flows control completely the experimental conditions. Also, these periodic flows can generate permanent convection flows by the mechanism of acoustic streaming [3] and of the so-called Schlichting mechanism. This overpasses strongly the scope of this paper, but it demonstrates that much attention has to be paid in the future to describe these sets of new forces and new phenomena which appear in micro-gravity conditions, due to vibrations.

In space for instance, experiments with liquids are flown in a rocket, in a shuttle or in the International Space Station (ISS). All these systems fly and run few experiments at a time, which all can generate some noise ; these experiments can also be perturbed by vibration generated by man activity. Anyway, as the global system is isolated, it preserves momenta. So, vibration induced by one experiment or by a man activity can perturb the other experiments. Due to this, the perturbation the experiment receives depends on (i) its own position $R_1$ in the ISS, but also on (ii) the position $R_2$ of the experiment that generates the noise, and (iii) of the polarisation $A_n$ of this noise.

For instance, if one assumes that the structure of the space shuttle or of the ISS is rigid, the movement of a man in one direction forces the ISS to move in the other direction ; hence, this generates a rotation $\Theta$ and the translation $T$ of the whole ISS itself, with a relative amplitude which depends on the direction $A_n$ of the man motion and of the relative distance $(R_1-R_o)/(R_2-R_o)$ to the mass centre $R_o$ of ISS. This motion perturbs the experiments.

This paper demonstrates, with the help of some example, how strongly an experiment can be perturbed by the modification of a boundary condition or of the vibration characteristics. It emphasises then the care one shall take to built up a spatial experiment on fluids and to chose its working location in the space station.

*Acknowledgements:* Discussion with E. Wesfreid, E. Guazelli, V. Kozlov, A. Ivanova, D. Lyubimov and T. Liubimova have been appreciated. CNES is thanked for partial funding.

# Feed-back from Readers :
## *Discussion, Comments and Answers*

**From *poudres & grains* articles:**

On *Poudres & grains* **12**, 17-42 (2001): In Figs 8,9,10, l3, Log means Neperian Logarithm. L/2 is the half of box length in Figs. 11, 12 & 13.

*From a remark by Y. Garrabos:* The rapid decrease of $V/(b\omega)$ *vs.* $\ln(2b/L)$, with the increase of b/L, is due to the synchronisation of the bead motion on the excitation, leading to regular impact at a frequency equal to half the excitation frequency $\nu$ {this occurs around $V/(b\omega)=4$ and $\ln(2b/L)=-2.53$, *i.e.* and corresponds to a one way ($\approx L$) which takes at time $1/\nu$} or at $\nu$ {this occurs around $V/(b\omega)=4$ and $\ln(2b/L)=-1.84$, *i.e.* round-trip ($\approx 2L$) takes $1/\nu$ }. This last synchronisation has been observed in recent Airbus result.

<div align="right">P.E.</div>

On *Poudres & grains* **12**, 60-82 (2001): *Remark by L. Ponson, P. Burban, H. Bellenger & P. Jean* : According to the data, $l_c=L/5$ to $L/6$ for the less dense sample.

***Answer:*** Indeed, the mean free path $l_c$ is smaller than the cell size L; however, it remains of the same order; so, this does not invalidate the qualitative findings and the theoretical discussion about the Knudsen regime; moreover, one expects that a decrease of $l_c/L$ generates an increase of loss, this value of $l_c/L$ may help understanding why the mean bead speed is so slow in the MniTexus experiment, compared to what one shall expect from 1-bead simulations.

<div align="right">P.E.</div>

On *Poudres & grains* **12**, 107-114 (2001): lines 6-7 of 2$^{nd}$ paragraph of section 3: the liquid flow exists but is small in the limit of a small toroidal section of radius $R_s$, *i.e.* proportional to $R_s/R_t$, due to preservation rules as stated in the introduction. In this limit this affects little the results. Furthermore, it is known that ripples formation occurs above the threshold at which the Stokes boundary layer is destabilised, *i.e.* when the Reynolds number $R_{e\delta}= b\Omega\delta/\nu=2b/\delta$ is larger than 100-to-500, for which $\delta= \sqrt{(2\nu/\Omega)}$ is the viscous boundary layer thickness and b is the relative amplitude of the flow motion; so, in the present case it scales as $R_{e\delta} =2\alpha_m R_s/\delta$ , for which $\alpha_m$ is the amplitude of rotation. It has been found also that ripple wavelength $\lambda$ is about $12\delta$ at threshold. Limit of Stability of the Stokes boundary layer can be found in V.G. Kozlov, *Stability of periodic motion of fluid in a planar* channel // Fluid Dynamics, 1979, vol. 14, no. 6, 904–908 and in Stability of high-frequency oscillating flow in channels // Heat Transfer – Soviet Research, 1991, v. 23, no. 7, 968–976.

<div align="right">P.E.</div>

## About published articles from other Reviews

About *Phys.Rev. Lett*. **84**, (2000) 5126, and *C.R. Physique* **3** (2002) 217-227, by J. Duran: Formation of such Ripples and its domain of existence has been studied under sinusoid vertical forcing by V.G. Kozlov, A. Ivanova and P. Evesque (Europhys. Lett. **42**,413-418 (1998)) with a viscous fluid instead of air. Similar heap formation has been found; it was also found .

<div align="right">P.E.</div>

The electronic arXiv.org version of this paper has been settled during a stay at the Kavli Institute of Theoretical Physics of the University of California at Santa Barbara (KITP-UCSB), in june 2005, supported in part by the National Science Fundation under Grant n° PHY99-07949.

*Poudres & Grains* can be found at :
http://www.mssmat.ecp.fr/rubrique.php3?id_rubrique=402